\documentclass[
preprint,showpacs,amsmath,amssymb,aps,pra
]{revtex4-1}

\usepackage{graphicx}
\usepackage{dcolumn}
\usepackage{bm}

\newcommand{\eref}[1]{(\ref{#1})}
\newcommand{\etal}{{\it et al.}}

\begin{document}

\title{Exact dynamics for optical coherent-state qubits subject to environment noise}

\author{Ming-Jay Yang}
\altaffiliation[Present address: ]{Institute of Photonics Technologies, National Tsing Hua University, Hsinchu 300, Taiwan}
\author{Shin-Tza Wu}
\email{phystw@gmail.com}
\affiliation{Department of Physics, National Chung Cheng University, Chiayi 621, Taiwan}

\date{\today}

\begin{abstract}
We study the exact dynamics of optical qubits encoded via coherent states with opposite phases which
are interacting with an environment modeled as a collection of simple harmonic oscillators. Making use
of a coherent-state path integral formulation, we are able to study memory effects on the dynamics of
the coherent-state qubits due to strong environment coupling. We apply this
formulation to examine the time evolution of a noisy quantum channel formed by two coherent-state qubits that are subject
to uncorrelated local environment noises. In particular, we examine the time evolution of
entanglement and maximal teleportation fidelity of the noisy quantum channel and show that at strong
coupling, due to large feedback effects from the environment noise, it is possible to maintain
a robust quantum channel in the long-time limit if appropriate error-correcting code is applied.
\end{abstract}


\maketitle

\section{\label{sec:int}Introduction}
Classical computation and information theories have chiefly been based on
encoding bit states for computing and information processing using
discrete classical variables taking two values, say, 0 and 1.
Quantum mechanics has opened up new possibilities for representing
the bit states making use of quantum mechanical states, such as
$|0\rangle$ and $|1\rangle$ along with their superpositions. This brings computational theory to a new
horizon, since new algorithms have proved to be able to solve problems that
are believed to be unsolvable classically \cite{Sho,Gro}. At the same time, it has also
revolutionized information theory in that new communication
protocols can reach unprecedented security level that is classically
impossible [\onlinecite{NC}--\onlinecite{Bou}].

In standard approaches to quantum computing and quantum information processing, one adopts two
orthogonal basis states, denoted $|0_L\rangle$ and $|1_L\rangle$ to encode the logical states of the
quantum bits (qubits), which can be two orthogonal spin states of electrons or nuclear moments,
or two orthogonal polarization states of single photons \cite{NC}. In recent years, however,
there has been a rapid growth of interest in an alternative approach which encodes quantum
information using continuous (quantum) variables \cite{cv1,cv2}. Since
unconditional quantum operations can be
achieved in this scheme, it has the merit of significantly reducing the resource
overhead for quantum information processing (though with non-ideal fidelities).
For optical implementations, the continuous-variable approach has the additional
advantage of experimental accessibility \cite{KL}. Typically, in optical implementations the
quantum information is encoded using the quadrature variables (for instance, the
``position" and the ``momentum" operators) of the electromagnetic fields which
have continuous spectra. The experimental detections of the quantum states can
then be achieved using homodyne detections with high efficiency and accuracy.
Moreover, various quantum optical techniques are available for quantum state
manipulations necessary for gate operations.

Along with these developments, Jeong and Kim \cite{Jeon02}, and Ralph \etal~\cite{Ralph03} propose to
encode the logical states of qubits using two coherent states with unequal amplitudes, for example,
\begin{eqnarray}
|0_L\rangle \mapsto |\alpha\rangle \qquad \mbox{and} \qquad |1_L\rangle \mapsto |\beta\rangle \, ,
\end{eqnarray}
where $|\alpha\rangle$, $|\beta\rangle$ are coherent states with (complex) amplitudes $\alpha$, $\beta$,
respectively, for an optical mode. The coherent states are defined as \cite{Walls}
\begin{eqnarray}
|\alpha\rangle \equiv e^{\alpha \hat{a}^\dagger - \alpha^* \hat{a}} |0\rangle
=e^{-\frac{|\alpha|^2}{2}} \sum_{n=0}^\infty \frac{\alpha^n}{\sqrt{n!}} |n\rangle \, ,
\label{CS}
\end{eqnarray}
where $\hat{a}$ is the annihilation operator for the optical mode, and $|n\rangle$ the number state
with $n$ photons, so that $|0\rangle$ represents the vacuum state.
Although coherent states with finite amplitudes are not exactly orthogonal to each other
(thus causing errors in quantum computing tasks and reduced fidelity in quantum information processing),
the coherent-state approach has several advantages. Among them, due to the continuous spectra of coherent states,
this scheme is thus a ``hybrid" of the discrete-variable and the continuous-variable approaches. It therefore
inherits merits from both approaches, so that unconditional single-qubit gate operations can be
implemented via offline resource states, linear optical networks, photon counting, and classical feedforward \cite{Ralph03}.
In particular, it has been shown that based on this scheme efficient quantum gates can be implemented \cite{Jeong} and
fault-tolerant quantum computation can be achieved with experimentally accessible amplitudes
for the coherent states \cite{Lund08}. At the same time, quantum error-correcting codes have also been developed for
the coherent-state logic [\onlinecite{Glancy}--\onlinecite{Wick13}]. Experimentally, the resource states for this scheme
(the ``cat states") can be generated using photon subtractions
[\onlinecite{Ourj07}--\onlinecite{Ourj09}], making the scheme a promising candidate for realistic quantum computing and
quantum information processing.

As with other realizations for quantum computing and quantum information processing,
coherent-state qubits are inevitably exposed to environment noises. For
instance, when an optical coherent-state passes through an optical element (e.g., a
beam splitter) that is part of a quantum gate, photon loss can occur due to finite absorption
coefficient of the element. In any realistic analysis, it is therefore essential to take into
account such effects. Earlier works in this regard have primarily focused on the limit of weak
qubit-environment interactions and/or negligible environment-noise coherence
time compared with the time scale for the qubit dynamics, so that the Born-Markov
approximation can be invoked \cite{Jeong,Glancy,Mun10,Wick10}.
In recent years, however, it has been recognized that such considerations are not
satisfactory for most realistic conditions. In particular, for full scale quantum computing it
may be necessary to integrate optical systems with, for instance, solid-state systems \cite{obrien}.
At the interface between these systems, more complicated decoherence mechanisms may arise
compared with those in all-optical settings. The study of non-Markovian dynamics
for open quantum systems has therefore become a key issue [\onlinecite{Wolf}--\onlinecite{LoF}].
In the present work, we aim to study the exact open-system dynamics for coherent-state
qubits making use of a formulation based on coherent-state
path integrals developed by Zhang and collaborators [\onlinecite{An}--\onlinecite{WM12}].
This formulation will allow us to study non-perturbatively the dynamics of coherent-state qubits
in the presence of strong environment noise. In particular, we will consider a noisy quantum channel, which consists of
two entangled qubits that are interacting with their local environments, and examine the time evolution of
its entanglement and teleportation fidelity. Surprisingly, we find that at strong coupling, due to
feedback effects from the environment noise, it is possible to preserve at long time the entanglement of
the two qubits and achieve better-than-classical teleportation fidelity if an appropriate error-correcting code
is applied. This demonstrates that it is feasible to establish a robust coherent-state quantum channel
even in the presence of environment noise.

We will start in Sec.~\ref{sec:form} by introducing a model for dissipation which allows exact
solutions via coherent-state path integral formulation. We will then examine the exact dynamics of a single
coherent-state qubit subject to such environment noise. In Sec.~\ref{sec:2qb} the analysis will be extended to
two entangled qubits that constitute a quantum channel. We will look into the time evolution of the entanglement
and teleportation fidelity of the quantum channel in the presence of dissipation.
Then in Sec.~\ref{sec:EC} we will study how error-correcting codes can help recover the entanglement and
teleportation fidelity of the pair of coherent-state qubits at long time. Finally, in Sec.~\ref{sec:fin} we
summarize our findings and offer brief discussions over related issues.

\section{\label{sec:form}Formulation}
To study decoherence of the coherent-state qubits, let us suppose the optical mode (henceforth the ``CS mode")
adopted for coherent-state encoding undergoes dissipation due to photon loss to its environment. This dissipation
has previously been modeled with a beam splitter which deflects photons from the CS mode into an environment mode
\cite{Glancy,Mun10,Wick10}. Since the dissipation is characterized solely by the
transmissivity of the beam splitter, it has no dynamics in this simple model \cite{Wick10}. Although one could phenomenologically
ascribe an exponentially decaying time-dependence to the transmissivity, the dynamics would invariably be Markovian
for which no memory effect from the environment coupling can arise \cite{Mun10}. In order to overcome this difficulty,
let us consider a generic model in which the CS mode interacts with an environment that consists of a collection of
simple harmonic oscillator modes. The total Hamiltonian thus reads (we set $\hbar=1$ throughout) \cite{loui}
\begin{eqnarray}
H = \omega_0 \hat{a}^\dagger \hat{a} + \sum_k \omega_k \hat{b}_k^\dagger \hat{b}_k
+ \sum_k \left( V_k \hat{a}^\dagger \hat{b}_k+ V_k^* \hat{b}_k^\dagger \hat{a} \right) \, ,
\label{H}
\end{eqnarray}
where $\omega_0$ is the CS mode frequency and $\hat{a}$ the corresponding annihilation operator,
$\hat{b}_k$ is the annihilation operator for the $k$-th environment mode with frequency $\omega_k$,
which is coupled to the CS mode with amplitude $V_k$. In this model, the CS mode exchanges
energy with each environment mode through a beam-splitter interaction Hamiltonian \cite{Weiss}.
These environment modes can correspond to, for instance, phonon modes in a solid or other photon modes.
The Hamiltonian \eref{H} therefore provides a generic model for photon loss which may be relevant for
interfacing between optical and solid-state systems, for instance in integrated quantum optical circuits
\cite{obrien}. In the limit of weak coupling, it reduces to a beam-splitter model with transmissivity that decays
exponentially with time (see later in this section) \cite{Jeong}. For general coupling, this generic model can exhibit richer dynamics
than that of the beam-splitter model, as we will see in the following \cite{Xion}. In particular, memory effects
due to environment feedback at strong coupling can lead to novel dynamics for the coherent-state qubits.

In the context of damped harmonic oscillators, the Hamiltonian \eref{H} has previously been studied
under the Born-Markov approximation \cite{loui,barnett}. In order to examine feedback
effects from the environment noise, it is necessary to go beyond this limit. This has been achieved by
Zhang and coworkers [\onlinecite{An}--\onlinecite{WM12}] using coherent-state path integrals applied to
the Feynman-Vernon influence functional formalism \cite{An,FV}. In essence, one starts from the
time evolution of the density matrix $\rho_{tot}$ for the total system (including the CS mode and the environment)
\begin{eqnarray}
\rho_{tot}(t) = e^{-iH(t-t_0)} \rho_{tot}(t_0) e^{+iH(t-t_0)} \, ,
\label{rho_tot}
\end{eqnarray}
where $t_0$ is the initial time and the Hamiltonian $H$ is given by \eref{H}. In the coherent-state representation,
one has the completeness relation \cite{Walls,note}
\begin{eqnarray}
\int \frac{d^2\alpha}{\pi} |\alpha\rangle\langle\alpha| = \hat{I} \, ,
\label{completeness}
\end{eqnarray}
where the integral extends over the entire complex $\alpha$-plane and
$d^2\alpha \equiv d\,\mbox{Re}\{\alpha\}\,d\,\mbox{Im}\{\alpha\}$ with Re,
Im indicating the real and imaginary parts, respectively; the coherent state
$|\alpha\rangle$ are defined earlier in \eref{CS}, and $\hat{I}$ is the identity operator.
Utilizing \eref{completeness}, one can express \eref{rho_tot} in terms of coherent-state
basis for the CS mode and the environment modes. By integrating out all environment degrees of freedom,
one can then arrive at an effective equation for the time evolution of the CS mode which is encoded with
the full environment effects.

Let us suppose the CS mode and the environment modes are completely decoupled initially, and the environment starts
off in the vacuum state at zero temperature. The initial total density matrix is thus
\begin{eqnarray}
\rho_{tot}(t_0) = \rho(t_0)\otimes|0_E \rangle \langle 0_E| \, ,
\label{rho_tot_t0}
\end{eqnarray}
where $\rho(t_0)$ is the initial density matrix for the CS mode and $|0_E\rangle$ denotes vacuum state for the
environment modes. Using \eref{rho_tot_t0} in \eref{rho_tot} and tracing out all environment modes in the
coherent-state basis, one can derive a path-integral representation for the time evolution of the reduced density matrix
for the CS mode \cite{An}. Expressing the reduced density matrix for the CS mode at time $t$
\begin{eqnarray}
\rho(t) = \int \frac{d^2\alpha_f}{\pi} \int \frac{d^2\alpha_f'}{\pi}
\rho(\alpha_f^*,\alpha_f';t) |\alpha_f\rangle\langle\alpha_f'| \, ,
\label{rho_t}
\end{eqnarray}
where $\alpha^*$ denotes the complex conjugate of $\alpha$ and
$\rho(\alpha_f^*,\alpha_f';t) \equiv \langle\alpha_f|\rho(t)|\alpha_f'\rangle$,
one finds in the influence functional formulation the time evolution for the matrix elements of the reduced
density matrix
\begin{eqnarray}
\rho(\alpha_f^*,\alpha_f';t)=\int \frac{d^2\alpha_i}{\pi} \int \frac{d^2\alpha_i'}{\pi}
\rho(\alpha_i^*,\alpha_i';t_0) {\cal K}(\alpha_i,\alpha_i',\alpha_f,\alpha_f'; t,t_0) \, ,
\label{rho_red}
\end{eqnarray}
where the kernel ${\cal K}$ now incorporates the entire environment effects in the Hamiltonian \eref{H} \cite{An,FV}.
At zero temperature the kernel is given by \cite{Xion}
\begin{eqnarray}
{\cal K}(\alpha_i,\alpha_i',\alpha_f,\alpha_f'; t,t_0) =
A(t) \exp\left\{\alpha_f^* u(t) \alpha_i + \alpha_i'^* B(t) \alpha_i + \alpha_i'^* u^*(t) \alpha_f' \right\}
\end{eqnarray}
with \cite{note}
\begin{eqnarray}
A(t) = e^{-\frac{1}{2} (|\alpha_i|^2 + |\alpha_i'|^2 +|\alpha_f|^2 +|\alpha_f'|^2 )}
\,,\quad
B(t) = 1-|u(t)|^2 \, .
\label{AB}
\end{eqnarray}
Here $u(t)$ follows the equation of motion
\begin{eqnarray}
\frac{d}{dt} u(t) + i\omega_0 u(t) + \int_{t_0}^t d\tau g(t-\tau) u(\tau) = 0 \,
\label{ut_eom}
\end{eqnarray}
subject to the initial condition $u(t_0)=1$. As we will notice in the following, $u(t)$ plays an
essential role in the dynamics of coherent-state qubits. In \eref{ut_eom} we have introduced the
noise correlation function
\begin{eqnarray}
g(t) = \int_0^\infty \frac{d\omega}{2\pi} J(\omega) e^{-i\omega t} \, ,
\label{gt}
\end{eqnarray}
where $J(\omega)$ is the spectral function for the CS mode-environment coupling in \eref{H}
\begin{eqnarray}
J(\omega) = \sum_k |V_k|^2 \delta(\omega - \omega_k) \, .
\end{eqnarray}
Namely it is the density of the environment modes weighted with the modulus square of the
coupling amplitude. For explicit calculations of the problem, one must have an explicit expression
for the spectral function. We will defer such calculations to later and focus for the moment
on establishing general formulations for the problems that will concern us.

Let us now apply the formulation above to study the decoherence dynamics of a single
coherent-state qubit. In the coherent-state encoding, the initial density matrix for a
single qubit has the general form
\begin{eqnarray}
\rho(t_0) = \sum_{m,n=1}^2 c_{mn} |\alpha_m\rangle\langle\alpha_n| \, ,
\label{rho_general}
\end{eqnarray}
where $c_{mn}$ are time-independent coefficients and $\alpha_{m,n}$ take values at the
encoding amplitudes. Throughout this work we will adopt coherent states with
opposite phases as the encoding basis. Therefore, for instance, with the encoding basis
$|\pm\alpha_0\rangle$, one would have in the equation above $\alpha_{1,2}=\pm\alpha_0$.
From \eref{rho_general}, it is clear that the time evolution of the density matrix
is entirely delegated to the elements $|\alpha_m\rangle\langle\alpha_n|$. Our first task
is therefore to work out the time evolution of such an element.

Let us consider an arbitrary element $\sigma(t_0)\equiv|\alpha\rangle\langle\beta|$. In the
coherent-state representation, its matrix elements are
\begin{eqnarray}
\sigma(\alpha_i^*,\alpha_i';t_0) &=& \langle\alpha_i|\Big(|\alpha\rangle\langle\beta|\Big)|\alpha_i'\rangle
= \langle\alpha_i|\alpha\rangle\langle\beta|\alpha_i'\rangle
\nonumber\\
&=& e^{-\frac{1}{2}\left(|\alpha_i|^2+|\alpha|^2-2\alpha_i^*\alpha\right)}
e^{-\frac{1}{2}\left(|\beta|^2+|\alpha_i'|^2-2\beta^*\alpha_i'\right)} \, .
\label{ab_mx}
\end{eqnarray}
The time evolution of the matrix element \eref{ab_mx} can be found using \eref{rho_red}, with
$\sigma$ here in place of the reduced density matrix $\rho$.
The integrals over $\alpha_i$ and $\alpha_i'$ are Gaussian integrals which can be dealt with easily
and yield
\begin{eqnarray}
\sigma(\alpha_f^*,\alpha_f';t)&=&e^{-\frac{1}{2}\left(|\alpha|^2+|\beta|^2\right)+\alpha(1-|u(t)|^2)\beta^*}
\nonumber\\
&\times& e^{-\frac{1}{2}\left(|\alpha_f|^2+|\alpha_f'|^2\right)+\alpha u(t) \alpha_f^*+\beta^*u^*(t)\alpha_f'} \, .
\label{ab_t}
\end{eqnarray}
Substituting \eref{ab_t} back into \eref{rho_t} and, as above, replacing $\rho$ with $\sigma$, one can carry out
the integrals over $\alpha_f$ and $\alpha_f'$, and arrive at the following prescription for the exact dynamics for the
element $|\alpha\rangle\langle\beta|$ in the presence of environment noise
\begin{eqnarray}
|\alpha\rangle\langle\beta| \longrightarrow
e^{-\frac{1-|u(t)|^2}{2} (|\alpha|^2+|\beta|^2-2\alpha\beta^*)}
|\alpha u(t)\rangle\langle\beta u(t)| \, ,
\label{elm}
\end{eqnarray}
where the arrow indicates time evolution. Equipped with \eref{elm}, we are now able to find the exact time evolution
for a coherent-state qubit with any initial density matrix. We will therefore refer to this result repeatedly in the
rest of this paper.

As an example, let us consider a coherent-state qubit initially in the ``cat state" in the basis
$\{|\pm\alpha_0\rangle\}$
\begin{eqnarray}
|Q\rangle = \frac{1}{\sqrt{N}} \left( c_1|\alpha_0\rangle + c_2|-\alpha_0\rangle \right) \, ,
\label{cat}
\end{eqnarray}
where $|c_1|^2+|c_2|^2=1$ and $N= 1 + e^{-2|\alpha_0|^2} (c_1^*c_2+c_1c_2^*)$ is a normalization
factor (note that it depends on $c_1$, $c_2$ and thus cannot be absorbed into them). Following the
prescription \eref{elm}, the time evolution of the state \eref{cat} subject to dissipations due to the
Hamiltonian \eref{H} can be obtained easily
\begin{eqnarray}
\rho(t) &=& \frac{1}{N} [\,|c_1|^2 |\alpha_t\rangle\langle\alpha_t| + |c_2|^2 |-\alpha_t\rangle\langle-\alpha_t|
\nonumber \\
&+& e^{-2(|\alpha_0|^2-|\alpha_t|^2)}
\left( c_1c_2^*|\alpha_t\rangle\langle-\alpha_t|+c_1^*c_2|-\alpha_t\rangle\langle\alpha_t|\right)\,] \, .
\label{rho_t_1}
\end{eqnarray}
Note that, for brevity here we have denoted
\begin{eqnarray}
\alpha_t \equiv \alpha_0 u(t) \, ,
\label{at}
\end{eqnarray}
which will also be used in the rest of this paper. Since the absolute value of $u(t)$ turns out to be always less than 1 for
$t>t_0$, the environment noise thus causes amplitude reduction in the coherent-state qubit \cite{note4} and induces
phase damping through the off-diagonal elements of the density matrix. In the limit of weak coupling, or when the coupling has a
broad spectrum, the spectral function $J(\omega)$ would have weak frequency dependence, so that the noise
correlation function \eref{gt} becomes a sharp function in time. Taking $t_0=0$, one can find from \eref{ut_eom} in this
limit \cite{barnett,Xion}
\begin{eqnarray}
u(t) \simeq e^{-\left( i \omega_0' + \frac{J(\omega_0)}{2} \right) t}
\label{markov}
\end{eqnarray}
with $\omega_0'=\omega_0+{\cal P}\int_0^\infty d\omega\frac{J(\omega)}{\omega-\omega_0}$, where $\cal P$ denotes
principal value of the integral. Therefore, in this limit the coherent-state amplitude decays exponentially
with time and Eq.~\eref{rho_t_1} reduces to the Markovian result obtained in Ref.~\onlinecite{Jeong}. For
general coupling strength, however, $u(t)$ can have quite different time dependence and novel
qubit dynamics can emerge, as we will soon discover.

It is interesting to note that the result \eref{rho_t_1} can be expressed in the form of an
operator sum \cite{Wick10,Kraus}
\begin{eqnarray}
\rho(t) = (1-p_e) |Q_t\rangle\langle Q_t| + p_e \hat{Z}|Q_t\rangle\langle Q_t|\hat{Z}^\dagger \, ,
\label{k_sum}
\end{eqnarray}
where $p_e\equiv (1-e^{-2(|\alpha_0|^2-|\alpha_t|^2)})/2$ and $|Q_t\rangle$ is the state $|Q\rangle$
of \eref{cat} with $|\pm\alpha_0\rangle$ replaced by $|\pm\alpha_t\rangle$ (the factor $N$ remains the same;
thus $|Q_t\rangle$ is not normalized), and we have introduced the ``Pauli-$Z$" operator $\hat{Z}$ such that
\begin{eqnarray}
\hat{Z} |\pm\alpha_t\rangle = \pm|\pm\alpha_t\rangle \, .
\label{pauliZ}
\end{eqnarray}
Note that here $\hat{Z}$ is neither Hermitian nor unitary \cite{note2}. It follows immediately from \eref{k_sum}
that the dynamical map induced by the environment noise consists of two parts: the mapping of
$|Q\rangle$ to $|Q_t\rangle$ (or damping of $\alpha_0$ to $\alpha_t$) and the random application of the Pauli-$Z$ operator.
We therefore recognize that the decoherence due to the interaction Hamiltonian in \eref{H} has a two-fold effect over
the coherent-state qubit \cite{Glancy}:
\begin{eqnarray}
&&\mbox{(a) reduction of the coherent-state amplitude through $u(t)$,} \nonumber\\
&&\mbox{(b) generation of random phase-errors with probability $p_e$.}
\label{effects}
\end{eqnarray}
As we shall find out, this is a crucial observation, since it suggests the appropriate error-correcting code
to be employed when one wishes to recover the coherence of the qubit \cite{Wick10}, which we shall discuss in Sec.~\ref{sec:EC}.

\section{\label{sec:2qb}Exact dynamics of two qubits}
Let us now turn to the problem of two coherent-state qubits under the influence of environment noise.
In this case, it would be interesting to look into a quantum channel formed by two entangled qubits
and examine how its quality degrades under the action of environment noise.
For this purpose, let us consider an initial state which has been of experimental
interest (the cluster-type entangled coherent state) \cite{Mun10,ct}
\begin{eqnarray}
|C\rangle =\frac{1}{\sqrt{M}} \left( |\alpha_0, \alpha_0 \rangle - z |\alpha_0, -\alpha_0 \rangle
- z |-\alpha_0, \alpha_0 \rangle - z^2 |-\alpha_0, -\alpha_0 \rangle\right) \, ,
\label{ctecs}
\end{eqnarray}
where $M=4(1 + e^{-4|\alpha_0|^2})$, $z=-i$ (which is kept implicit here
for later convenience), and
$|\alpha_0,\alpha_0\rangle=|\alpha_0\rangle\otimes|\alpha_0\rangle$ etc
denote coherent states of the two CS modes in question. Here, again, each qubit is
encoded with the $\{|\pm\alpha_0\rangle\}$ basis states. When the two CS modes
are subject to independent dissipations induced by the Hamiltonian \eref{H} with the
same spectral function, the time evolution of the CS modes can then be found in accordance
with the single qubit case. Namely, with each mode following the prescription \eref{elm},
we have the following for the time evolution of any element in a two-qubit density matrix
\begin{eqnarray}
&&|\alpha,\alpha'\rangle\langle\beta,\beta'|=|\alpha\rangle\langle\beta|\otimes|\alpha'\rangle\langle\beta'|
\nonumber\\
&&\rightarrow
e^{-\frac{1-|u(t)|^2}{2} (|\alpha|^2+|\alpha'|^2+|\beta|^2+|\beta'|^2-2(\alpha\beta^*+\alpha'\beta'^*))}
|\alpha u(t), \alpha' u(t)\rangle\langle\beta u(t), \beta' u(t)| \, .
\end{eqnarray}
It is then easy to work out the exact dynamics for the initial state \eref{ctecs}.
Its density matrix at any time $t>t_0$ is found to be
\begin{eqnarray}
&&\rho(t) = \frac{1}{M} [\,
|\alpha_t,\alpha_t\rangle\langle\alpha_t,\alpha_t|
-ic|\alpha_t,\alpha_t\rangle\langle\alpha_t,-\alpha_t|
-ic|\alpha_t,\alpha_t\rangle\langle-\alpha_t,\alpha_t|
+c^2|\alpha_t,\alpha_t\rangle\langle-\alpha_t,-\alpha_t|
\nonumber\\
&&+ic|\alpha_t,-\alpha_t\rangle\langle\alpha_t,\alpha_t|
+|\alpha_t,-\alpha_t\rangle\langle\alpha_t,-\alpha_t|
+c^2|\alpha_t,-\alpha_t\rangle\langle-\alpha_t,\alpha_t|
+ic|\alpha_t,-\alpha_t\rangle\langle-\alpha_t,-\alpha_t|
\nonumber\\
&&+ic|-\alpha_t,\alpha_t\rangle\langle\alpha_t,\alpha_t|
+c^2|-\alpha_t,\alpha_t\rangle\langle\alpha_t,-\alpha_t|
+|-\alpha_t,\alpha_t\rangle\langle-\alpha_t,\alpha_t|
+ic|-\alpha_t,\alpha_t\rangle\langle-\alpha_t,-\alpha_t|
\nonumber\\
&&+c^2|-\alpha_t,-\alpha_t\rangle\langle\alpha_t,\alpha_t|
-ic|-\alpha_t,-\alpha_t\rangle\langle\alpha_t,-\alpha_t|
-ic|-\alpha_t,-\alpha_t\rangle\langle-\alpha_t,\alpha_t|
+|-\alpha_t,-\alpha_t\rangle\langle-\alpha_t,-\alpha_t|\,] \, .
\nonumber\\ &&
\label{rho_t_ctecs}
\end{eqnarray}
Here $M$ is the same as in \eref{ctecs} and we have denoted
$c\equiv 1-2p_e=\exp\{-2(|\alpha_0|^2-|\alpha_t|^2)\}$.
Note that this result can also be obtained using the operator sum formulation
by extending \eref{k_sum} to two CS modes subject to
independent, identical dissipations \cite{wu12}. Equation \eref{rho_t_ctecs}
thus describes the exact time evolution of a noisy quantum channel initially
in the state \eref{ctecs} with each CS mode under the action of the Hamiltonian \eref{H}.

To examine the quality of the noisy quantum channel, we shall study the change
of its entanglement property and teleportation ability with time. To this end, we shall calculate
the time evolution of the concurrence \cite{Woot} and maximal teleportation fidelity \cite{Horod}
for the density matrix \eref{rho_t_ctecs}. For bipartite two-state systems, the concurrence for
a density matrix $\rho$ is defined as
\begin{eqnarray}
C \equiv \max\{0,\sqrt{\lambda_1}-\sqrt{\lambda_2}-\sqrt{\lambda_3}-\sqrt{\lambda_4}\} \,,
\label{C_def}
\end{eqnarray}
where $\lambda_i$ are eigenvalues of $\rho(\hat{Y}_1 \otimes \hat{Y}_2)\rho^*(\hat{Y}_1 \otimes \hat{Y}_2)$ with
$\lambda_1$ being the largest one. Here $\hat{Y}_j$ are Pauli-$Y$ operators for sub-system $j$ and $\rho^*$ is
the complex conjugate of the density matrix $\rho$ \cite{Woot}. To find the concurrence for $\rho(t)$ of \eref{rho_t_ctecs},
it is thus necessary to have first a matrix representation for the density matrix with respect to an orthonormal basis
for the restricted two-mode space spanned by $\{|\pm\alpha_t\rangle\otimes|\pm\alpha_t\rangle\}$.
Let us consider the following ``even" and ``odd" states
\begin{eqnarray}
|e\rangle = \frac{1}{\sqrt{N_e}} \left( |\alpha_t\rangle + |-\alpha_t\rangle \right)
\quad \mbox{and} \quad
|o\rangle = \frac{1}{\sqrt{N_o}} \left( |\alpha_t\rangle - |-\alpha_t\rangle \right) \, ,
\label{eo}
\end{eqnarray}
where $N_{e,o}=2(1\pm e^{-2|\alpha_t|^2})$ are normalization factors. One can easily check using \eref{CS} that
$|e\rangle$ and $|o\rangle$ consist of number states with, respectively, even
and odd numbers of photons, and hence are orthogonal to each other. We can therefore
employ the basis set $\{|ee\rangle,|eo\rangle,|oe\rangle,|oo\rangle\}$ for the two CS modes
and arrive at the following matrix representations for the original basis states
\begin{eqnarray}
|\alpha_t,\pm\alpha_t\rangle = \left(
                                 \begin{array}{c}
                                      a^2 \\
                                   \pm ab \\
                                       ab \\
                                   \pm b^2 \\
                                 \end{array}
                               \right)
\, , \qquad
|-\alpha_t,\pm\alpha_t\rangle = \left(
                                 \begin{array}{c}
                                      a^2 \\
                                   \pm ab \\
                                      -ab \\
                                   \mp b^2 \\
                                 \end{array}
                               \right) \, .
\label{basis}
\end{eqnarray}
Here we have denoted
\begin{eqnarray}
a = \sqrt{\frac{1+e^{-2|\alpha_t|^2}}{2}}
\quad \mbox{and} \quad
b = \sqrt{\frac{1-e^{-2|\alpha_t|^2}}{2}} \, .
\label{ab}
\end{eqnarray}
Using \eref{basis} in \eref{rho_t_ctecs}, we arrive at the matrix representation
for the density matrix in the basis $\{|ee\rangle,|eo\rangle,|oe\rangle,|oo\rangle\}$
\begin{eqnarray}
\rho(t) = \frac{4}{M}
          \left(
            \begin{array}{cccc}
              a^4 (1+c^2) & 0 & 0 & 2ic a^2 b^2 \\
              0 & a^2 b^2(1-c^2) & 0 & 0 \\
              0 & 0 & a^2 b^2 (1-c^2) & 0 \\
              -2ic a^2 b^2 & 0 & 0 & b^4 (1+c^2) \\
            \end{array}
          \right) \, .
\label{rho_t_oe}
\end{eqnarray}
We note that the density matrix takes an ``X-form" \cite{YE_X}. Its concurrence
can thus be found relatively easily. We get
\begin{eqnarray}
C=\frac{2 a^2 b^2}{1+e^{-4|\alpha_0|^2}} \max\{0, c^2+2c-1\} \, .
\label{conc}
\end{eqnarray}

The maximal teleportation fidelity (henceforth ``teleportation fidelity" for short) for a quantum channel
with density matrix $\rho$ is defined through its fully entangled fraction
\begin{eqnarray}
f_{\rm max} \equiv \max_{|\psi\rangle} \langle\psi|\rho|\psi\rangle \, ,
\label{fmax_def}
\end{eqnarray}
where the maximum is taken over all possible maximally entangled states $|\psi\rangle$. For two-state systems,
the teleportation fidelity is then given by \cite{Horod}
\begin{eqnarray}
F=\frac{2 f_{\rm max} + 1}{3} \, .
\label{Fmax}
\end{eqnarray}
To find the fully entangled fraction for the noisy channel \eref{rho_t_ctecs}, one can first recast the
density matrix in terms of the following orthonormal basis set
\begin{eqnarray}
|\phi_1\rangle &\equiv&  |\Phi^+\rangle = \frac{1}{\sqrt{2}} \left( |ee\rangle + |oo\rangle \right)
\, ,\nonumber\\
|\phi_2\rangle &\equiv& i|\Phi^-\rangle = \frac{i}{\sqrt{2}} \left( |ee\rangle - |oo\rangle \right)
\, ,\nonumber\\
|\phi_3\rangle &\equiv& i|\Psi^+\rangle = \frac{i}{\sqrt{2}} \left( |eo\rangle + |oe\rangle \right)
\, ,\nonumber\\
|\phi_4\rangle &\equiv&  |\Psi^-\rangle = \frac{1}{\sqrt{2}} \left( |eo\rangle - |oe\rangle \right)
\,,
\end{eqnarray}
where $|\Phi^\pm\rangle$, $|\Psi^\pm\rangle$ are the usual Bell states in the even-odd basis.
The fully entangled fraction $f_{\rm max}$ then corresponds to the largest eigenvalue for the real part
of the transformed density matrix \cite{bennett}. This calculation leads to
\begin{eqnarray}
f_{\rm max} = \frac{1}{2(1+e^{-4|\alpha_0|^2})} (c^2 - 2 a^2 b^2 (1-c)^2 +1) \, .
\label{fmax}
\end{eqnarray}
Because here each qubit is restricted to a two-dimensional state space, the teleportation fidelity for
the noisy channel \eref{rho_t_ctecs} can thus be obtained by substituting \eref{fmax} back into \eref{Fmax}.

\begin{figure}
\begin{center}
\includegraphics*[width=80mm]{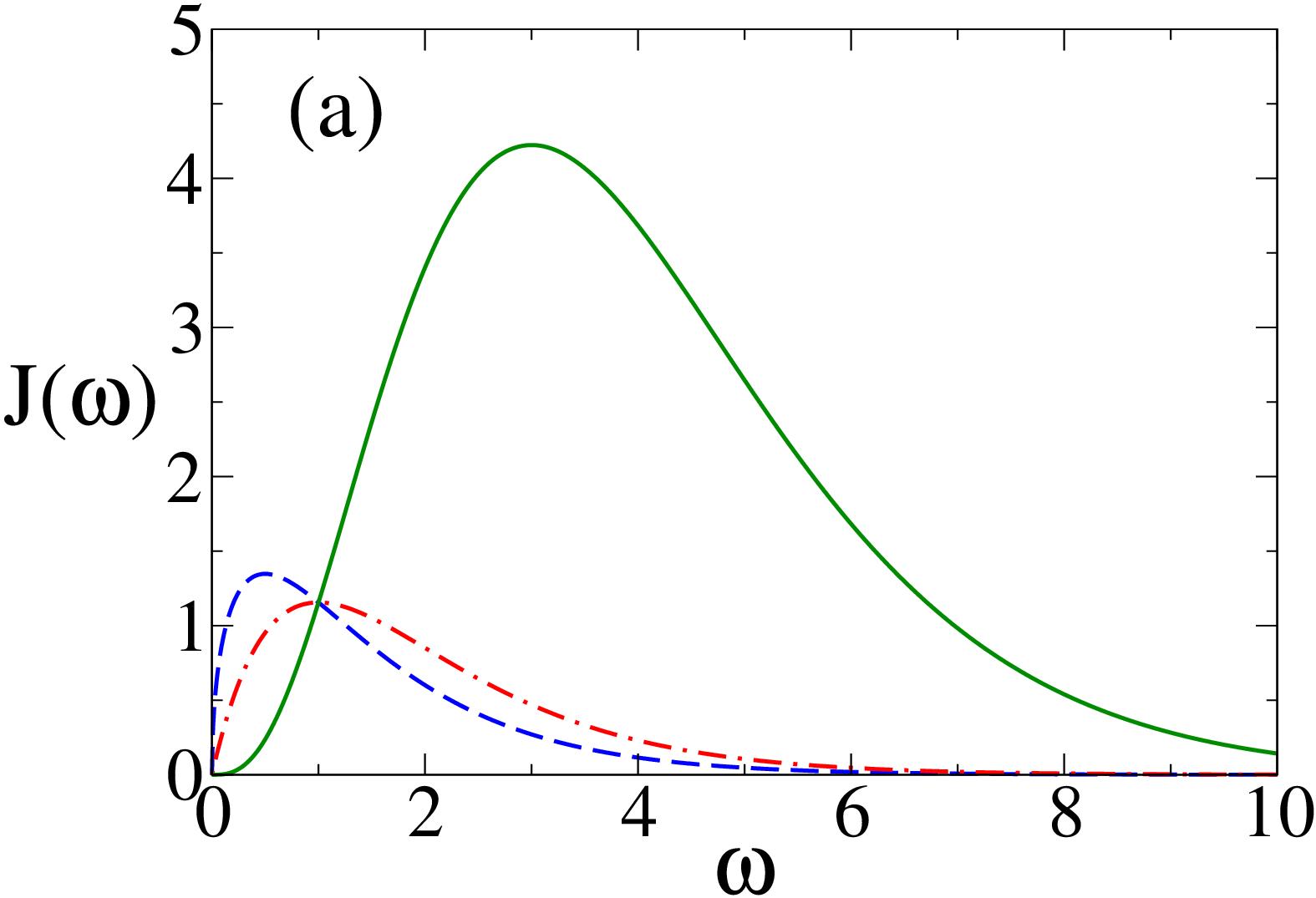}
\includegraphics*[width=80mm]{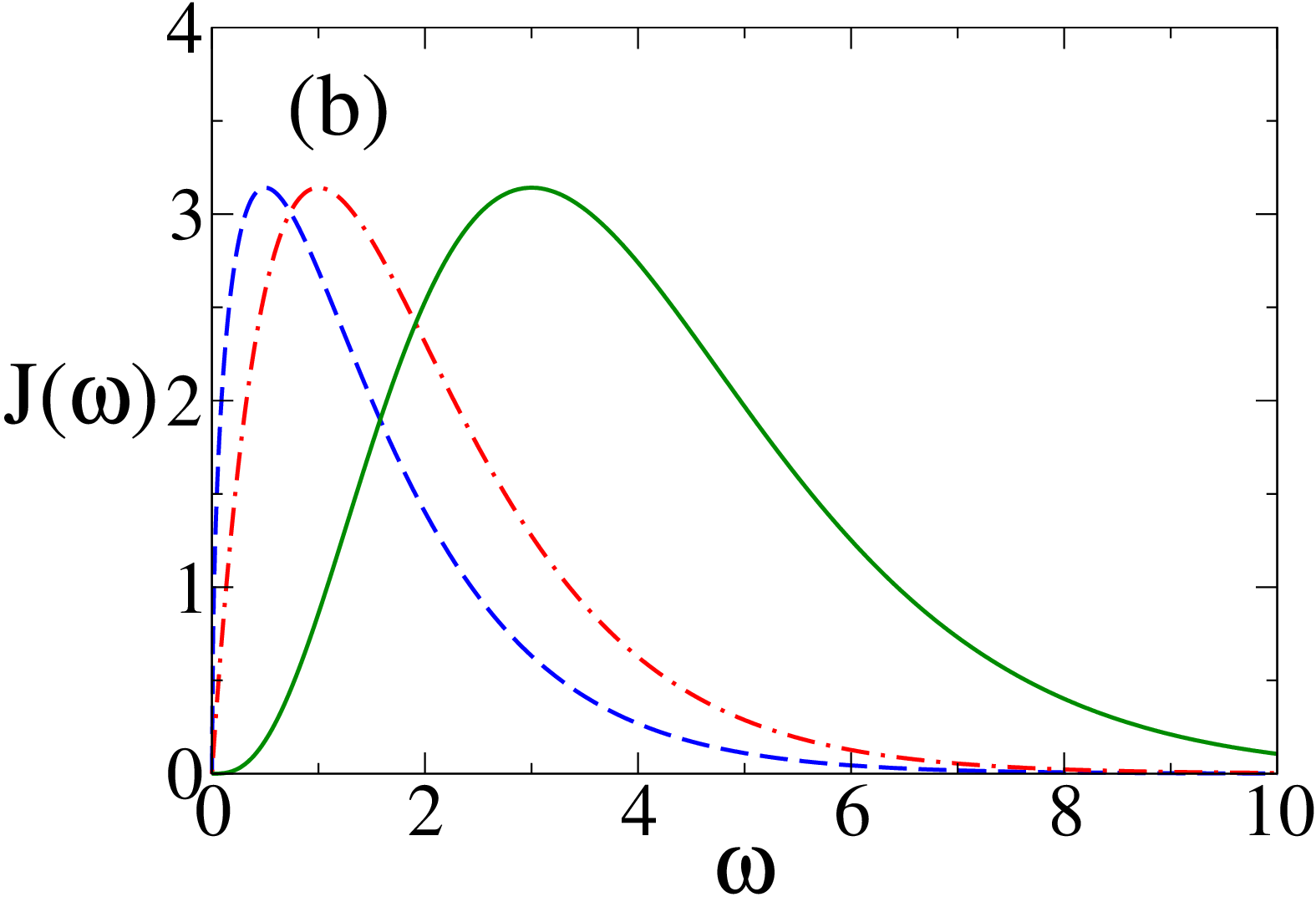}
\end{center}
\caption{Spectral function $J(\omega)$ for $\eta_0=0.5$ (a) without scaling (i.e. $\eta_s=\eta_0$ in \eref{Jw})
and (b) with scaling (i.e.~with $\eta_s$ given by \eref{eta_s})
for sub-Ohmic ($s=1/2\,$; blue dashed curves),
Ohmic ($s=1\,$; red dot-dashed curves), and super-Ohmic ($s=3\,$; green solid curves) cases.
Here the horizontal axes are plotted in units of $\omega_c$, which is taken to be 1 in all plots throughout this paper.
\label{fig:Jw}}
\end{figure}

We shall now proceed with explicit calculations for noise models for the results above. This requires a specific form
for the spectral function $J(\omega)$. Here we will consider the following form of a power law
with an exponential cutoff \cite{Weiss}
\begin{eqnarray}
J(\omega) = 2\pi\eta_s\omega\left(\frac{\omega}{\omega_c}\right)^{s-1}
\exp\left(-\frac{\omega}{\omega_c}\right) \, ,
\label{Jw}
\end{eqnarray}
where (cf.~Ref.~\onlinecite{Weiss})
\begin{eqnarray}
\eta_s=\eta_0 \left(\frac{e}{s}\right)^s \, .
\label{eta_s}
\end{eqnarray}
Here $\eta_0$ is the coupling strength and $\omega_c$ the cutoff frequency, which is much larger than
any other frequency scales in the problem. It is common to categorize the spectral function \eref{Jw}
into three classes according to the power $s$ of the frequency variable $\omega$: the sub-Ohmic ($0<s<1$), Ohmic ($s=1$), and
super Ohmic ($s>1$) ones \cite{Weiss}. Note that unlike earlier literatures, here we have defined $\eta_s$ in the form
\eref{eta_s} so that for the same $\eta_0$ and $\omega_c$, the spectral function \eref{Jw} would have
the same peak height $2\pi\eta_0\omega_c$ for all $s>0$; Figure \ref{fig:Jw} illustrates a comparison for
$J(\omega)$ without scaling and with scaling. This would make the meaning of
``coupling strength" less ambiguous when one compares results for $J(\omega)$ with different power $s$.
This is because for given $s$ the peak position of $J(\omega)$ occurs at $\omega=s\omega_c$, its coupling
is therefore ``detuned" by $s\omega_c-\omega_0$ from the CS mode frequency $\omega_0$. Now that $J(\omega)$ for different $s$
have the same peak height (for the same $\eta_0$ and $\omega_c$), by comparing the ``detunning", one can have
a clear picture as to which power $s$ would generate stronger coupling for the CS mode.

\begin{figure}
\begin{center}
\includegraphics*[width=80mm]{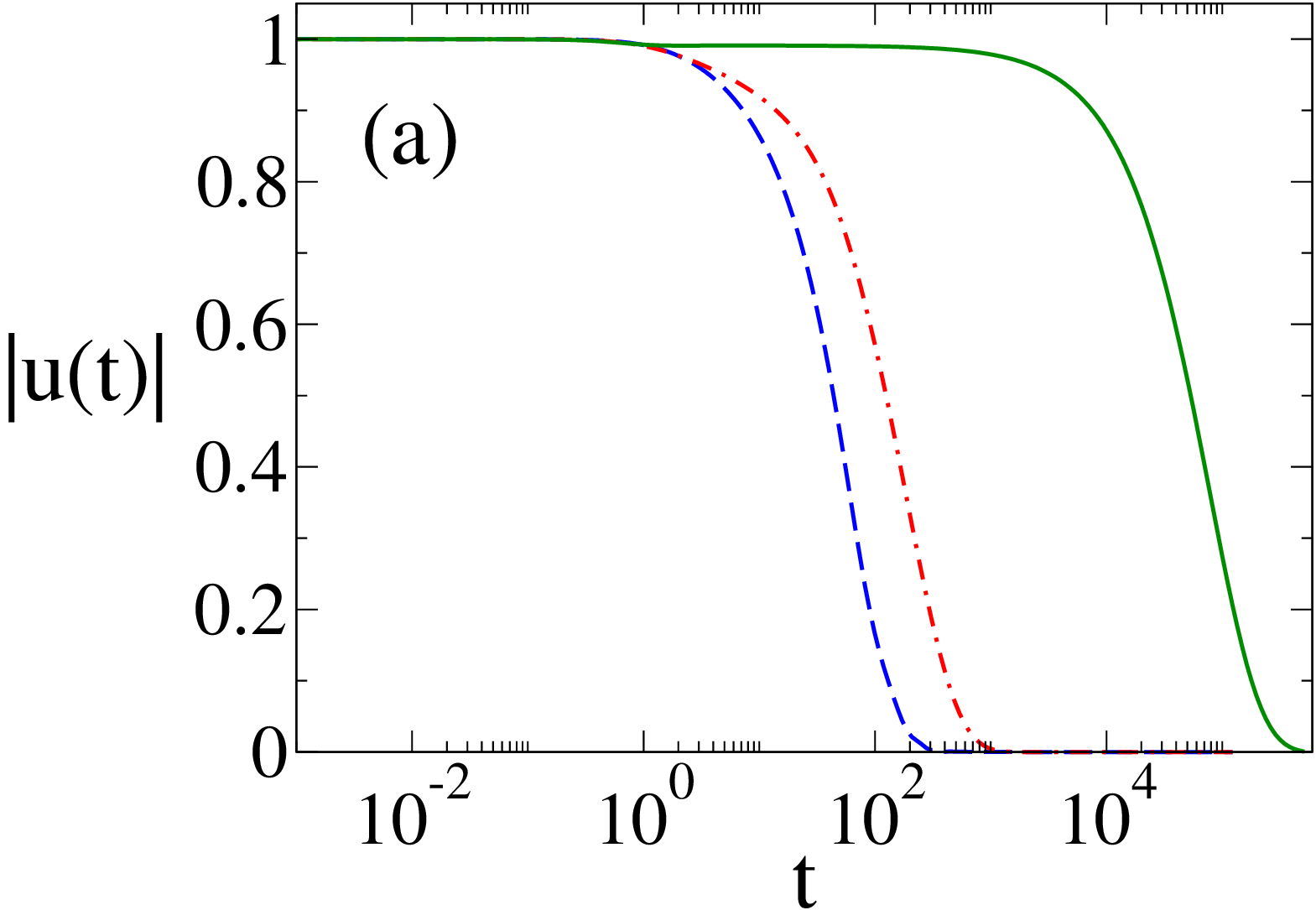}
\includegraphics*[width=80mm]{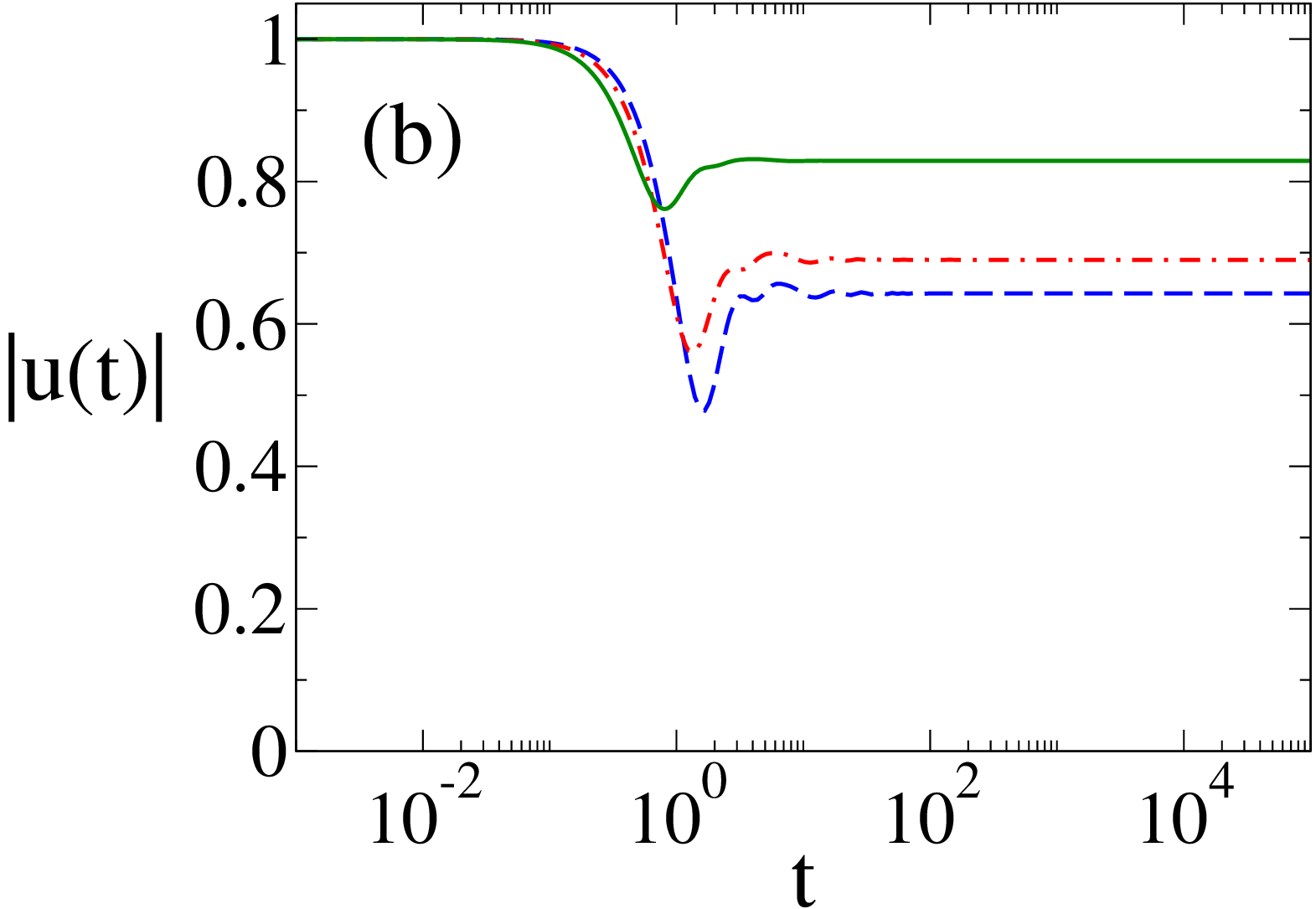}
\end{center}
\caption{The absolute value of $u(t)$ for coupling strength (a) $\eta_0=0.01$ and (b) 0.5 for sub-Ohmic
($s=1/2\,$; blue dashed curves), Ohmic ($s=1\,$; red dot-dashed curves), and super-Ohmic
($s=3\,$; green solid curves) cases. Note that here and in all subsequent plots, we take $t_0=0$ and
use log scale for the time axis, which is in units of $1/\omega_c$.
\label{fig:ut}}
\end{figure}

For the spectral function \eref{Jw}, it is not possible to obtain analytic solution for $u(t)$ from
the equation of motion \eref{ut_eom}. Using techniques of Laplace transformation, however, one can
express $u(t)$ as Bromwich integrals involving special functions (such as exponential integrals
and error functions). Besides contribution from contours around branch-cut for the integrand,
depending on the coupling strength $\eta_0$, the integral can also receive contribution
from poles of the integrand. When the pole contribution exists, $u(t)$ would tend to
non-zero steady value at long time \cite{WM12}. We relegate details of these calculations
to Appendix \ref{sec:ut}. Figure \ref{fig:ut} illustrates our results for $|u(t)|$ for the
coupling strengths $\eta_0=0.01$ and $\eta_0=0.5$ for different power $s$ in the spectral
function \eref{Jw}. Here, following Ref.~\onlinecite{Xion} we consider $s=1/2$ for
sub-Ohmic coupling, and $s=3$ for super-Ohmic coupling. We note that at weak coupling ($\eta_0=0.01$),
$|u(t)|$ decays exponentially with time, while at strong coupling ($\eta_0=0.5$),
after a sharp drop initially it recovers gradually and eventually saturates at non-zero
value at long time. The weak coupling result (Fig.~\ref{fig:ut}(a)) exhibits typical Markovian dynamics
\cite{note3}. For the strong coupling result (Fig.~\ref{fig:ut}(b)),
the sharp decay is due to the large coupling between the qubit and the environment modes,
which leads to a stronger decay initially than that at weak coupling. However, feedback
from the environment modes subsequently bring $|u(t)|$ back and a long-time correlation
between the qubit and the environment modes is gradually established, leading to
non-dissipative $|u(t)|$ evolution at long time \cite{Xion}. In view of Eq.~\eref{effects},
this means that at strong coupling the amplitude decay of the qubit saturates in the long-time limit.
A natural question thus arises: For the noisy quantum channel
\eref{rho_t_ctecs}, would we have a robust, non-dissipative quantum channel at long time?
Namely, would the concurrence and the teleportation fidelity of the noisy quantum channel possess,
respectively, non-zero steady values and better than classical values in the long-time limit?

\begin{figure*}
\includegraphics*[width=120mm]{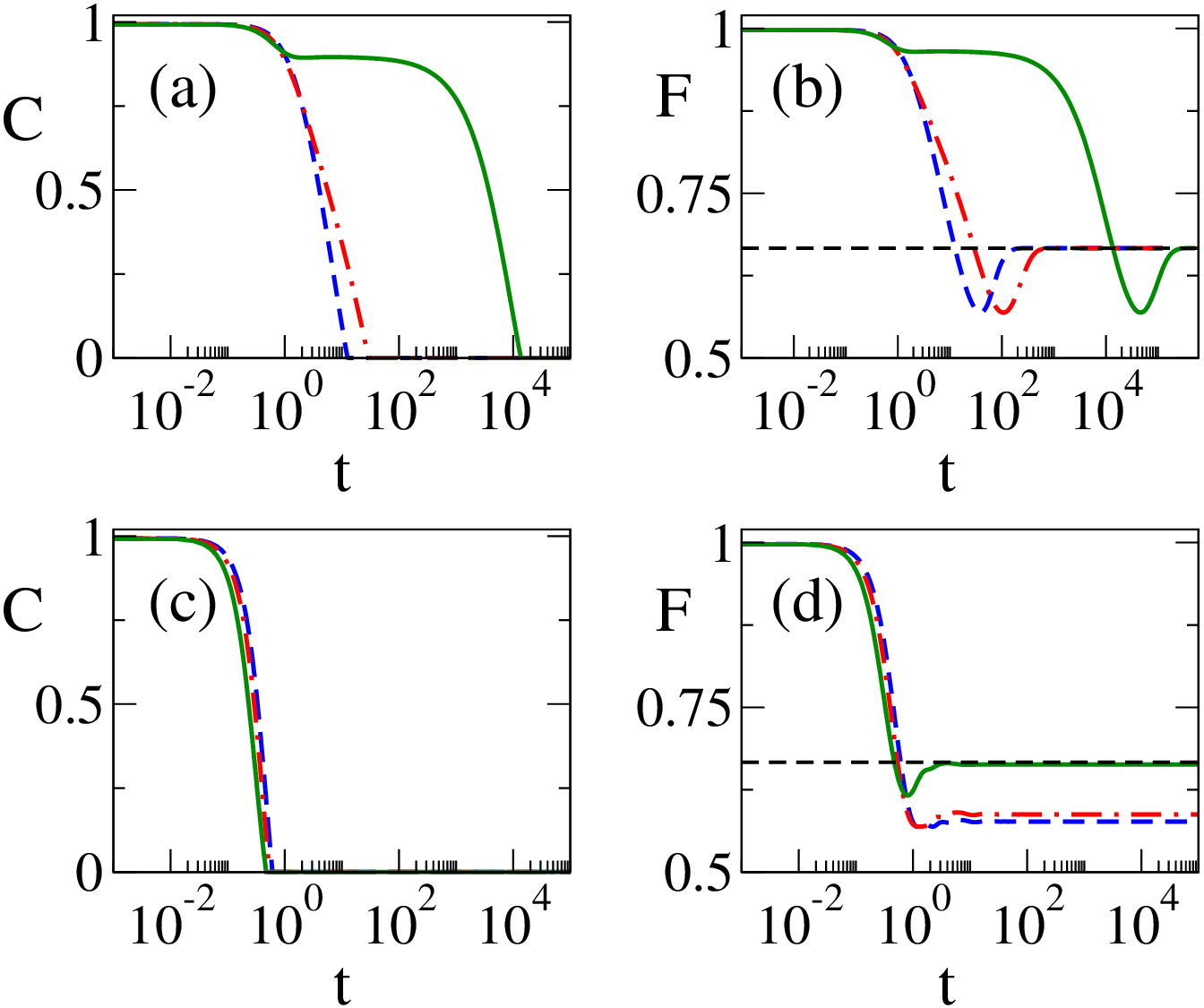}
\caption{Concurrence $C$ and teleportation fidelity $F$ for the noisy quantum channel \eref{rho_t_ctecs}
at weak coupling ($\eta_0=0.01$; panels (a) and (b)) and strong coupling ($\eta_0=0.5$; panels (c) and (d))
for sub-Ohmic ($s=1/2\,$; blue dashed curves), Ohmic ($s=1\,$; red dot-dashed curves), and
super-Ohmic ($s=3\,$; green solid curves) cases. The horizontal dashed lines in (b) and (d) indicate the
classical limit $F=2/3$ for the teleportation fidelity. In all panels, the time axes are plotted in units
of $1/\omega_c$.
\label{fig:CtFt}}
\end{figure*}

To answer the questions above, we supply the numerically obtained $u(t)$ to formulas for the concurrence
\eref{conc} and the teleportation fidelity \eref{Fmax} (by way of \eref{fmax}) for the noisy quantum channel
\eref{rho_t_ctecs}. For the CS mode, we shall consider an experimentally realistic initial amplitude $\alpha_0=1.2$
[\onlinecite{Ourj07}--\onlinecite{Ourj09}] and the frequency $\omega_0=0.1\omega_c$, since the cutoff
frequency $\omega_c$ is the largest frequency scale in the problem.
The results for our calculation are demonstrated in Fig.~\ref{fig:CtFt} for weak ($\eta_0=0.01$) and strong
($\eta_0=0.5$) couplings with sub-Ohmic ($s=1/2$), Ohmic, and super-Ohmic
($s=3$) spectral functions. From Fig.~\ref{fig:CtFt}(a), we see that at weak coupling the concurrence decays monotonically
to zero for all three cases, with the sub-Ohmic case having the largest decay rate and the super-Ohmic
one the smallest. This can be understood easily because the super-Ohmic case has the largest detunning
(with its peak at $\omega=3\omega_c$) from the CS mode frequency ($\omega_0=0.1\omega_c$), while the
sub-Ohmic case has the smallest (with peak position at $\omega=0.5\omega_c$). In the long-time limit,
as Fig.~\ref{fig:CtFt}(b) shows,
the teleportation fidelities of all three cases drop to the classical value $2/3\simeq0.667$ \cite{pope}. Thus,
as anticipated, at weak coupling the quality of the noisy channel \eref{rho_t_ctecs} does degrade with time.

For strong coupling, despite the non-dissipative evolution of $u(t)$ at long time, we see in
Figs.~\ref{fig:CtFt}(c), (d) that the concurrence decays monotonically to zero at even faster rates than at weak
coupling, and the teleportation fidelity even falls below the classical value at long time. In other words,
in terms of the two figures-of-merit considered here, at strong coupling the quantum channel has a lower
quality than that at weak coupling. This is surprising because the long-time correlation in $u(t)$ turns out
not helpful in preserving the quantum channel at long time, in spite of the non-dissipative time evolution.
Nevertheless, if one recalls from Eq.~\eref{effects} that the effects of the environment noise on the qubits
are in fact two-fold, it is then clear why such results emerge: The non-dissipative evolution at long time
is not enough to support the quantum channel because effect (b) in \eref{effects} is still in action here.
Namely, it is the random phase-errors between the two qubits that disrupt the quantum channel at strong coupling.
Therefore, in order to sustain a robust quantum channel at long time, we need to take care of both effects
in \eref{effects} properly. This is the task we shall now turn to.

\section{\label{sec:EC}Exact dynamics of two qubits with error corrections}
In order to combat against effect (b) in \eref{effects} due to the environment noise,
we shall resort to schemes of quantum error-corrections. Now that we have identified these errors
being due to random phase-flips, it is natural to adopt phase-flip error-correcting codes
\cite{Glancy,Wick10}. In Sec.~\ref{sec:pfc}, we will therefore examine the exact dynamics
of the noisy quantum channel when a phase-flip error-correcting code is applied.
As a comparison, in Sec.~\ref{sec:bfc} we will also consider an encoding scheme
(the ``bit-flip encoding") that was previously proposed for constructing quantum channels using
coherent-state qubits \cite{Mun10}.

\subsection{\label{sec:pfc}Phase-flip error correction}
To correct the random phase-flip error, we shall employ a scheme proposed by
Glancy \etal~\cite{Glancy} for coherent-state qubits. Take three-bit encoding as an example, in this
scheme the signal qubit is sent at the encoding stage together with two ancilla modes in vacuum
states from the sender's side. Making use of two sets of beam splitters simulating two CNOT operations,
one subsequently applies three Hadamard gates so that the qubits can be protected against phase-flip
errors. Here the Hadamard gate operates in the way that (up to normalization factors) \cite{note_H}
\begin{eqnarray}
|\pm\alpha\rangle \xrightarrow{\text{\tiny Hadamard}} |\alpha\rangle \pm |-\alpha\rangle \,
\label{Hadamard}
\end{eqnarray}
with $|\pm\alpha\rangle$ the basis states for the qubit.
After passing through the noisy region, the encoded qubit first passes through another three
Hadamard gates for decoding and then a sequence of beam-splitters and photodetectors for
syndrome detection. According to the error syndromes, one can apply corrective operations to
recover the signal qubit at the receiving end. In this way, one would be able to correct one phase-flip error
in the qubits \cite{NC}. More errors can be corrected similarly when more encoding ancilla modes are
added. In general, an $n$-bit encoding in the present scheme can correct up to $(n-1)/2$ phase-errors
($n$ must be odd number) and the probability for an error-free transmission is \cite{NC,Glancy}
\begin{eqnarray}
p_s = \sum_{k=0}^{\frac{n-1}{2}}
\left(
\begin{array}{c}
           n \\ k
\end{array}
\right)
(1-p_e)^{n-k} p_e^k \, ,
\label{ps}
\end{eqnarray}
where $p_e$ is the probability for one phase-flip error in each mode. Note that when $p_e < 1/2$,
the success probability $p_s$ can be made arbitrarily close to 1 with sufficiently large $n$.

To incorporate the error-correcting scheme above into our calculation for the exact dynamics of
the noisy quantum channel \eref{rho_t_ctecs}, we note that the probability $p_e$ for
phase-flip errors in our calculation is given by that in \eref{k_sum}.
Thus, according to \eref{ps}, with an $n$-bit error-correcting code implemented the error probability becomes
$p_e' = 1-p_s$. This corresponds to changing the previously defined $c$ in \eref{rho_t_ctecs} into
\begin{eqnarray}
c' = 1 - 2 p_e' = 2 p_s -1 \, .
\end{eqnarray}
The time evolution for the concurrence and the fully entangled fraction for the error-corrected noisy
quantum channel can now be obtained from \eref{conc} and \eref{fmax} by simply replacing $c$ with $c'$
above.

\begin{figure*}
\includegraphics*[width=120mm]{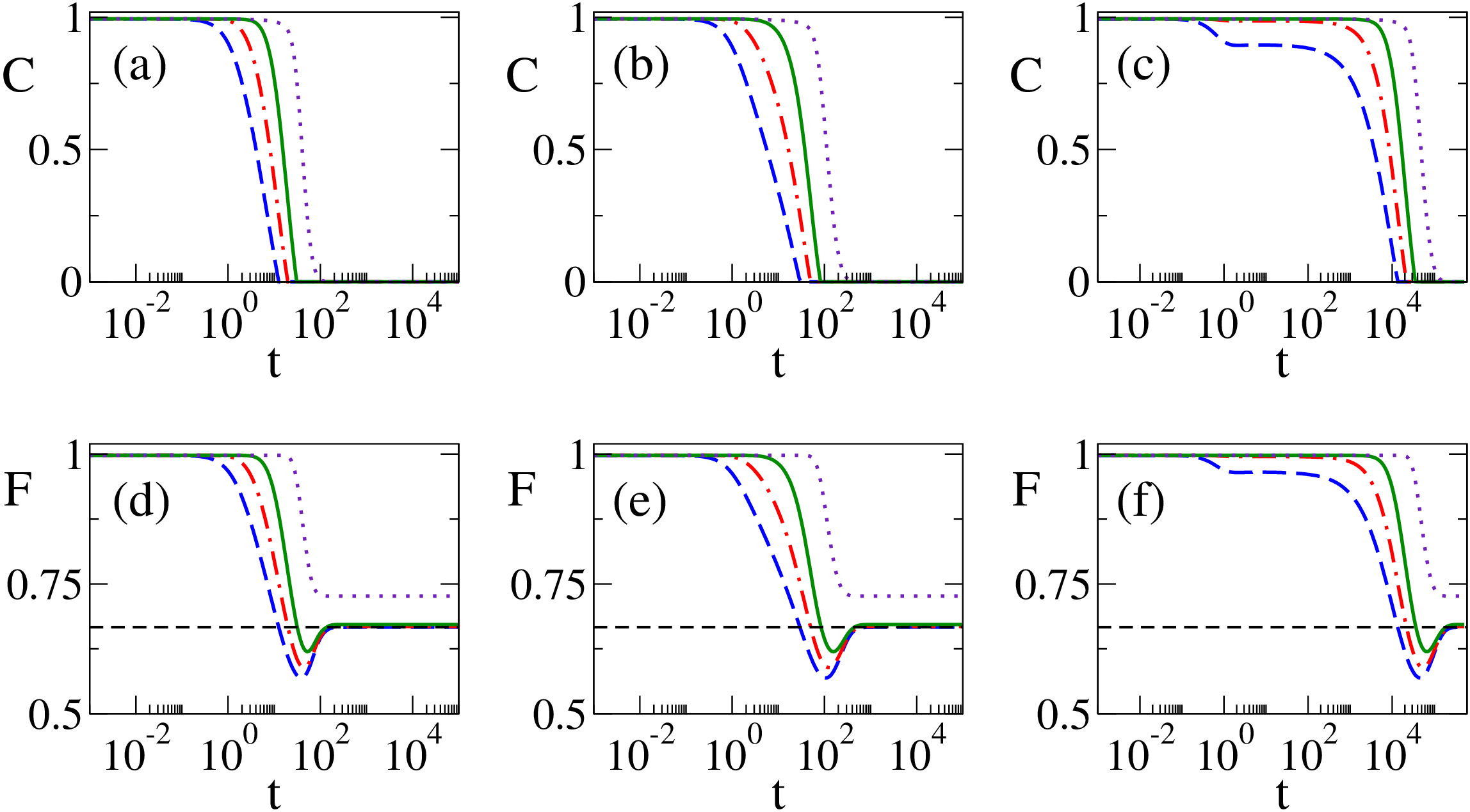}
\caption{Concurrence $C$ and teleportation fidelity $F$ for the noisy channel at weak coupling ($\eta_0=0.01$) for
sub-Ohmic ($s=1/2\,$; panels (a), (d)),
Ohmic ($s=1\,$; panels (b), (e)), and
super-Ohmic ($s=3\,$; panels (c), (f)) cases without encoding (blue dashed curves), and with
phase-flip encoding using 3-bit (red dot-dashed curves), 9-bit (green solid surves), and 101-bit (purple dotted curves) codes.
In each of panels (d)--(f), the horizontal dashed line signifies the classical limit $F=2/3$.
\label{fig:pf:CtFt001}}
\end{figure*}

\begin{figure*}
\begin{center}
\includegraphics*[width=120mm]{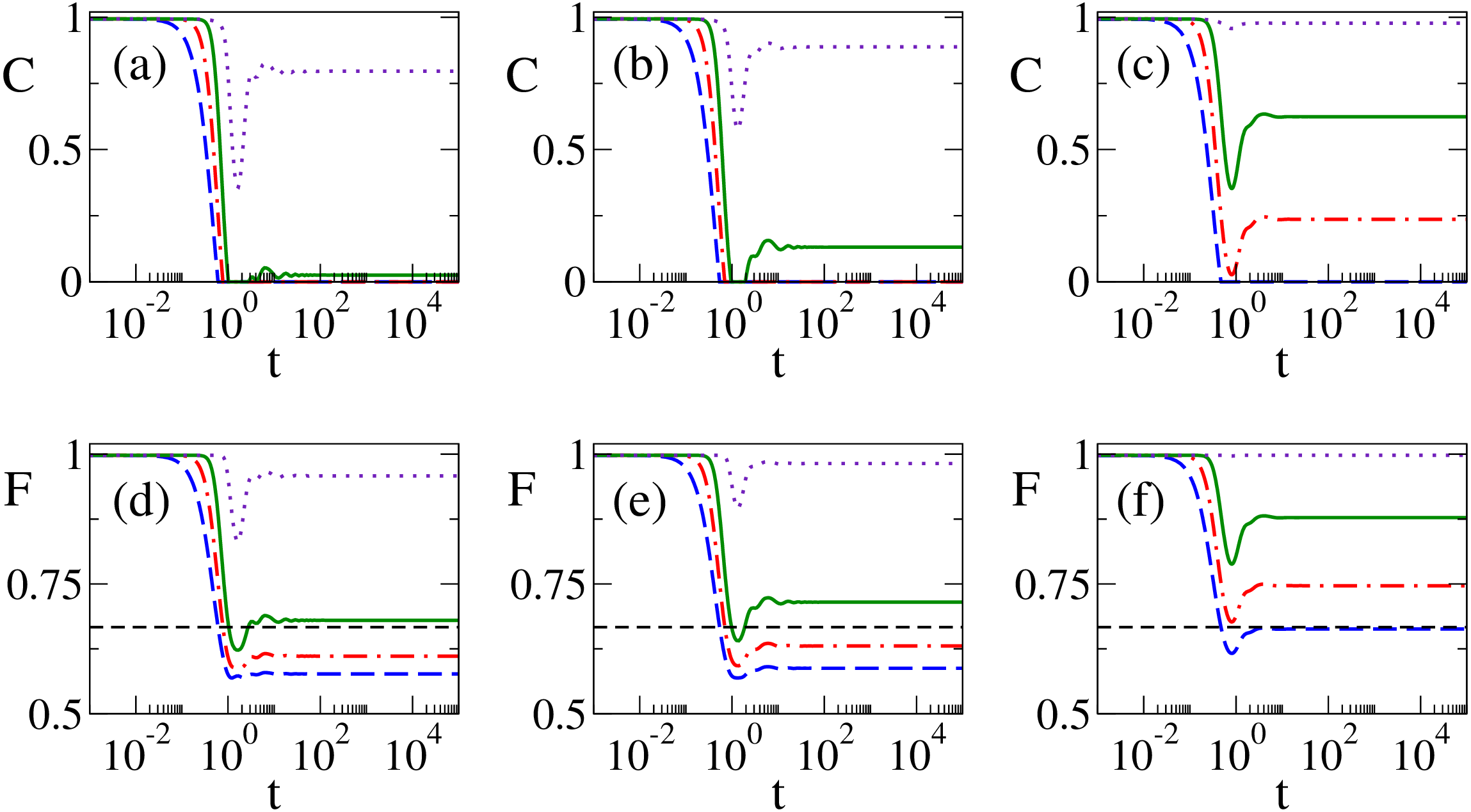}
\end{center}
\caption{The same as Fig.~\ref{fig:pf:CtFt001} but for strong coupling ($\eta_0=0.5$).
\label{fig:pf:CtFt05}}
\end{figure*}

The results for these calculations are shown in Fig.~\ref{fig:pf:CtFt001}
for weak coupling ($\eta_0=0.01$) and Fig.~\ref{fig:pf:CtFt05} for strong coupling ($\eta_0=0.5$).
One can notice immediately that at weak coupling the concurrence for all three cases have now attained longer life spans,
and the teleportation fidelity can go above the classical limit after error correction is applied.
These gains, however, do not seem to be very impressive as even with a 101-bit encoding, the enhancement in
the teleportation fidelity is still rather limited ($F\simeq 0.727$ at large time for all three cases). Such limited gain
at so high cost does not seem practical.

At strong coupling, however, the situation turns out quite different. We see from Fig.~\ref{fig:pf:CtFt05}
that in this case, error correction can improve the concurrence and the
teleportation fidelity quite significantly. For super-Ohmic coupling, with 3-bit encoding the noisy channel
can maintain a non-zero concurrence ($\simeq 0.237$; see Fig.~\ref{fig:pf:CtFt05}(c)) and a better than classical
teleportation fidelity ($\simeq 0.747$; see Fig.~\ref{fig:pf:CtFt05}(f)) at long time.
With 9-bit encoding, we find that the concurrence for all three classes of spectral functions attain finite steady values
and the teleportation fidelity are all raised above the classical value at long time.
In the limit of large $n$ encoding (illustrated with $n=101$ here), even for the sub-Ohmic
case (which would induce the strongest environment coupling due to its small detunning), the concurrence can be recovered
to 0.796, and the teleportation fidelity to 0.958. Therefore, at strong coupling, the phase-flip error-correcting code
can recover the noisy channel significantly and result in a robust quantum channel at long time.

To understand the results above, let us recall from \eref{effects} the two effects (a) amplitude reduction, and (b) random
phase-errors due to the environment noise. As noted above, for error probability $p_e<1/2$ the present error-correcting code can
correct phase-flip errors very efficiently. For coherent-state qubits, from \eref{k_sum}, since $p_e<1/2$ always,
the error-correcting code can protect the noisy channel against effect (b) very well.
The difference between the weak-coupling and the strong-coupling results, therefore, is primarily due to effect (a).
At weak coupling, since $|u(t)|$ decays with time monotonically to zero (see Fig.~\ref{fig:ut}(a)),
effect (a) persists all the way until the qubit is completely damped away. At strong coupling, however, $|u(t)|$
saturates at long time and thus effect (a)
is entirely removed when this steady is reached. Therefore, for strong coupling, when the phase-flip error-correcting
code is implemented, both effects from the environment noise can be accounted for and a robust quantum channel
can persist at long time.

\subsection{\label{sec:bfc}Bit-flip encoding}
In the preceding subsection, we have seen that when $u(t)$ becomes non-dissipative in the long-time limit,
applying phase-flip error-correcting code can help preserve the quantum channel. As a comparison, here we shall
consider a different encoding scheme for coherent-state qubits. In Ref.~\onlinecite{Mun10}, a repetition encoding
was proposed for the quantum channel \eref{ctecs} and its performance under photon loss
has been analyzed in the Markovian limit. With exact dynamics for the coherent-state qubits available, here
we revisit this problem in particular for strong environment coupling when non-dissipative dynamics of $u(t)$ is present.
Since this encoding scheme is identical to that in bit-flip error-correcting codes \cite{NC},
in the following we will refer to it as the ``bit-flip encoding" (note that no error correction is attempted in this scheme \cite{Mun10}).
Following Ref.~\onlinecite{Mun10}, an $n$-bit encoding in this scheme yields for the state \eref{ctecs}
\begin{eqnarray}
|C_n\rangle =\frac{1}{\sqrt{M_n}} &&\left(
|\alpha_0\rangle^{\otimes n} |\alpha_0 \rangle^{\otimes n}
- z^n |\alpha_0\rangle^{\otimes n} |-\alpha_0 \rangle^{\otimes n}
\right.\nonumber\\ && \left.
- z^n |-\alpha_0\rangle^{\otimes n} |\alpha_0 \rangle^{\otimes n} |
- z^{2n} |-\alpha_0\rangle^{\otimes n} |-\alpha_0 \rangle^{\otimes n} \right) \, ,
\label{ctecs_n}
\end{eqnarray}
where the normalization factor
\begin{eqnarray}
M_n = \left\{
\begin{array}{ll}
4                         & \mbox{for even $n$,} \\
4 (1+e^{-4n|\alpha_0|^2}) & \mbox{for odd $n$,}
\end{array}
\right.
\end{eqnarray}
and as in \eref{ctecs} $z=-i$; here
$|\alpha\rangle^{\otimes n}\equiv|\alpha\rangle\otimes\dots\otimes|\alpha\rangle$ etc
denote direct products of $n$ coherent states. Since each basis state $|\pm\alpha_0 \rangle^{\otimes n}$
consists of $n$ independent modes, the time evolution for an element of the density matrix for
the encoded state \eref{ctecs_n} at zero temperature can be generalized straightforwardly from \eref{elm}
\begin{eqnarray}
|\alpha\rangle^{\otimes n}\langle\beta|^{\otimes n} \rightarrow
e^{-\frac{1-|u(t)|^2}{2} (|\alpha|^2+|\beta|^2-2\alpha\beta^*)n}
|\alpha u(t)\rangle^{\otimes n}\langle\beta u(t)|^{\otimes n} \,
\label{elm_n}
\end{eqnarray}
with $u(t)$, as before, determined from \eref{ut_eom}. The time evolution of the initial density matrix
$|C_n\rangle\langle C_n|$ for \eref{ctecs_n} then follows easily from the prescription \eref{elm_n}
\begin{eqnarray}
&&\rho(t) = \frac{1}{M_n} [\,
|\alpha_t\rangle^{\otimes n}|\alpha_t\rangle^{\otimes n}\langle\alpha_t|^{\otimes n}\langle\alpha_t|^{\otimes n}
-z^{*n}c^n|\alpha_t\rangle^{\otimes n}|\alpha_t\rangle^{\otimes n}\langle\alpha_t|^{\otimes n}\langle-\alpha_t|^{\otimes n}
\nonumber\\
&&-z^{*n}c^n|\alpha_t\rangle^{\otimes n}|\alpha_t\rangle^{\otimes n}\langle-\alpha_t|^{\otimes n}\langle\alpha_t|^{\otimes n}
-z^{* 2n}c^{2n}|\alpha_t\rangle^{\otimes n}|\alpha_t\rangle^{\otimes n}\langle-\alpha_t|^{\otimes n}\langle-\alpha_t|^{\otimes n}
\nonumber\\
&&-z^n c^n|\alpha_t\rangle^{\otimes n}|-\alpha_t\rangle^{\otimes n}\langle\alpha_t|^{\otimes n}\langle\alpha_t|^{\otimes n}
+|\alpha_t\rangle^{\otimes n}|-\alpha_t\rangle^{\otimes n}\langle\alpha_t|^{\otimes n}\langle-\alpha_t|^{\otimes n}
\nonumber\\
&&+c^{2n}|\alpha_t\rangle^{\otimes n}|-\alpha_t\rangle^{\otimes n}\langle-\alpha_t|^{\otimes n}\langle\alpha_t|^{\otimes n}
+z^{* n} c^n|\alpha_t\rangle^{\otimes n}|-\alpha_t\rangle^{\otimes n}\langle-\alpha_t|^{\otimes n}\langle-\alpha_t|^{\otimes n}
\nonumber\\
&&-z^n c^n|-\alpha_t\rangle^{\otimes n}|\alpha_t\rangle^{\otimes n}\langle\alpha_t|^{\otimes n}\langle\alpha_t|^{\otimes n}
+c^{2n}|-\alpha_t\rangle^{\otimes n}|\alpha_t\rangle^{\otimes n}\langle\alpha_t|^{\otimes n}\langle-\alpha_t|^{\otimes n}
\nonumber\\
&&+|-\alpha_t\rangle^{\otimes n}|\alpha_t\rangle^{\otimes n}\langle-\alpha_t|^{\otimes n}\langle\alpha_t|^{\otimes n}
+z^{* n} c^n|-\alpha_t\rangle^{\otimes n}|\alpha_t\rangle^{\otimes n}\langle-\alpha_t|^{\otimes n}\langle-\alpha_t|^{\otimes n}
\nonumber\\
&&-z^{2n} c^{2n}|-\alpha_t\rangle^{\otimes n}|-\alpha_t\rangle^{\otimes n}\langle\alpha_t|^{\otimes n}\langle\alpha_t|^{\otimes n}
+z^n c^n|-\alpha_t\rangle^{\otimes n}|-\alpha_t\rangle^{\otimes n}\langle\alpha_t|^{\otimes n}\langle-\alpha_t|^{\otimes n}
\nonumber\\
&& +z^n c^n|-\alpha_t\rangle^{\otimes n}|-\alpha_t\rangle^{\otimes n}\langle-\alpha_t|^{\otimes n}\langle\alpha_t|^{\otimes n}
+|-\alpha_t\rangle^{\otimes n}|-\alpha_t\rangle^{\otimes n}\langle-\alpha_t|^{\otimes n}\langle-\alpha_t|^{\otimes n} ] \, ,
\nonumber\\ &&
\label{rho_t_ctecs_n}
\end{eqnarray}
where $\alpha_t$ is, as previously, given by \eref{at} and $c$ the same as in \eref{rho_t_ctecs}. To find the concurrence and
teleportation fidelity for the noisy channel \eref{rho_t_ctecs_n}, again one has to express $\rho(t)$ in terms of orthonormal
basis sets. Similar to \eref{eo}, we adopt the $n$-bit repetition even-odd states
\begin{eqnarray}
|e_n\rangle = \frac{1}{\sqrt{M_e}} \left( |\alpha_t\rangle^{\otimes n} + |-\alpha_t\rangle^{\otimes n} \right)
\quad \mbox{and} \quad
|o_n\rangle = \frac{1}{\sqrt{M_o}} \left( |\alpha_t\rangle^{\otimes n} - |-\alpha_t\rangle^{\otimes n} \right) \, ,
\label{eo_n}
\end{eqnarray}
where  $M_{e,o}=2(1\pm e^{-2n|\alpha_t|^2})$. The calculation then proceeds in close parallel with that for the non-encoded case
in Sec.~\ref{sec:2qb}, except that care must be taken over the distinction between $n$ being even or odd. In the basis
$\{|e_n e_n\rangle,|e_n o_n\rangle,|o_n e_n\rangle,|o_n o_n\rangle\}$, one finds for even $n$
\begin{eqnarray}
\rho(t) =\left(
            \begin{array}{cccc}
              a_n^4 & -i^n a_n^3 b_n c^n &  -i^n a_n^3 b_n c^n & -a_n^2 b_n^2 c^{2n} \\
               -i^n a_n^3 b_n c^n & a_n^2 b_n^2 & a_n^2 b_n^2 c^{2n} & i^n a_n b_n^3 c^n\\
               -i^n a_n^3 b_n c^n & a_n^2 b_n^2 c^{2n} & a_n^2 b_n^2 & i^n a_n b_n^3 c^n\\
             -a_n^2 b_n^2 c^{2n} & i^n a_n b_n^3 c^n & i^n a_n b_n^3 c^n & b_n^4 \\
            \end{array}
          \right) \,,
\label{rho_t_ctecs_even}
\end{eqnarray}
and for odd $n$
\begin{eqnarray}
\rho(t) = \frac{4}{M_n}
          \left(
            \begin{array}{cccc}
              a_n^4 (1+c^{2n}) & 0 & 0 & 2 i^n a_n^2 b_n^2 c^n \\
              0 & a_n^2 b_n^2(1-c^{2n}) & 0 & 0 \\
              0 & 0 & a_n^2 b_n^2(1-c^{2n}) & 0 \\
              -2 i^n a_n^2 b_n^2 c^n & 0 & 0 & b_n^4 (1+c^{2n}) \\
            \end{array}
          \right) \, .
\label{rho_t_ctecs_odd}
\end{eqnarray}
In both \eref{rho_t_ctecs_even} and \eref{rho_t_ctecs_odd}, we
have denoted
\begin{eqnarray}
a_n = \sqrt{\frac{1 + e^{-2n|\alpha_t|^2}}{2}} \quad \mbox{and} \quad
b_n = \sqrt{\frac{1 - e^{-2n|\alpha_t|^2}}{2}} \, .
\label{anbn}
\end{eqnarray}
Accordingly, in the same manners as before, one can find for the bit-flip encoded noisy channel the concurrence
\begin{eqnarray}
C=\frac{8 a_n^2 b_n^2}{M_n} \max\{0, c^{2n}+2c^n-1\}
\label{conc_n}
\end{eqnarray}
and the fully entangled fraction
\begin{eqnarray}
f_{\rm max} =
\left\{
   \begin{array}{cc} \displaystyle
          \frac{1}{4} \left( 1+4 a_n^2 b_n^2 c^{2n} +\sqrt{(a_n^2-b_n^2)^4 + 16 a_n^2 b_n^2 c^{2n}} \right) \,
          & \mbox{for even $n$,} \\ \displaystyle
          \frac{1}{2(1+e^{-4n|\alpha_0|^2})} (c^{2n} - 2 a_n^2 b_n^2 (1-c^n)^2 +1) \,
          & \mbox{for odd $n$.}
   \end{array}
\right.
\label{fmax_n}
\end{eqnarray}
As usual, using \eref{fmax_n} in \eref{Fmax}, one can obtain the teleportation fidelity for the encoded noisy channel.

\begin{figure*}
\includegraphics*[width=120mm]{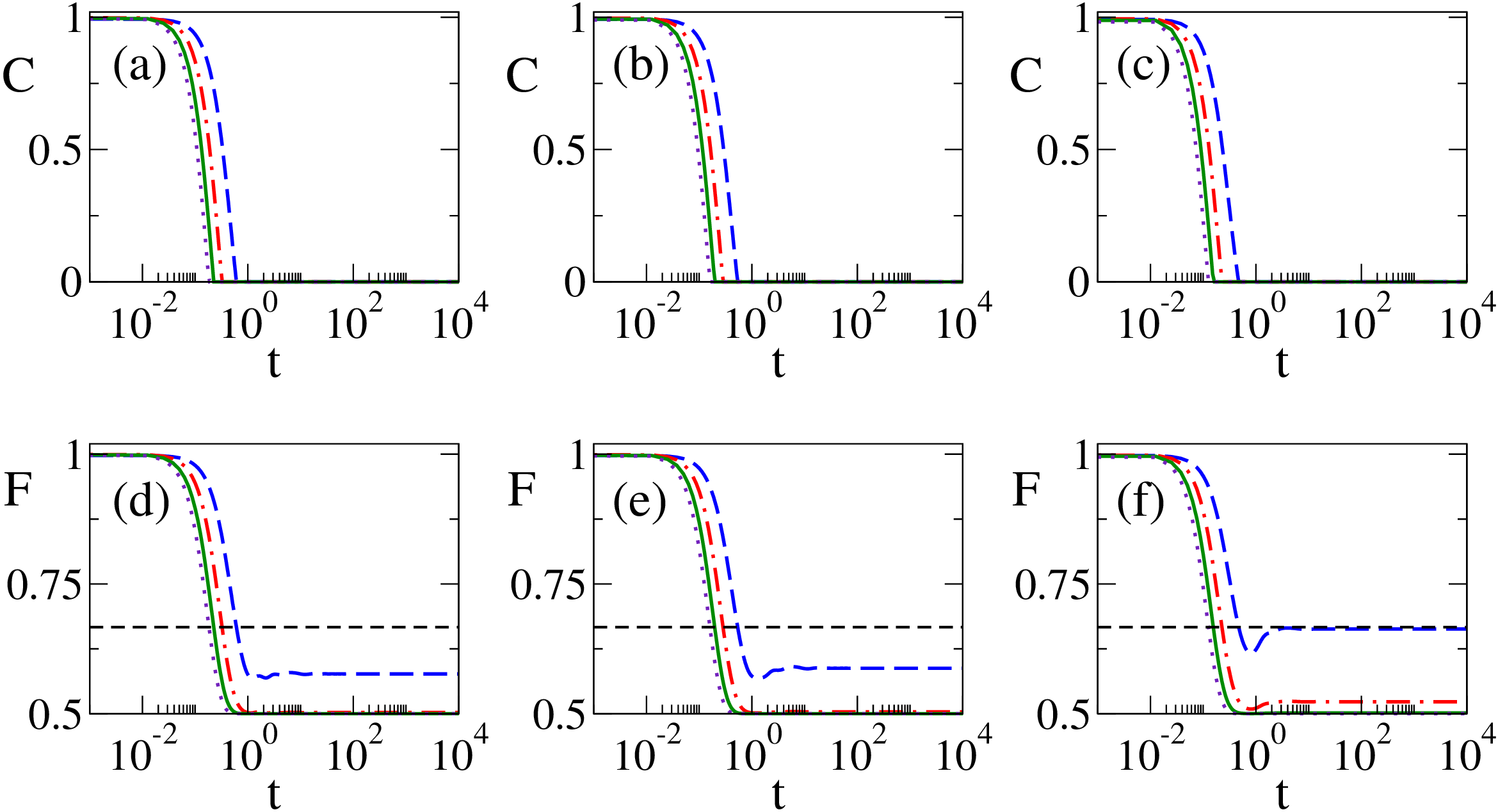}
\caption{Concurrence $C$ and teleportation fidelity $F$ for the noisy channel at strong coupling ($\eta_0=0.5$) for
sub-Ohmic ($s=1/2\,$; panels (a), (d)),
Ohmic ($s=1\,$; panels (b), (e)), and
super-Ohmic ($s=3\,$; panels (c), (f)) cases without encoding (blue dashed curves), and with
bit-flip encoding using 3-bit (red dot-dashed curves), 6-bit (green solid curves), and 9-bit (purple dotted curves) codes.
As in previous figures, the horizontal dashed lines in (d)--(f) indicate the classical limit $F=2/3$.
\label{fig:bf:CtFt05}}
\end{figure*}

Figure \ref{fig:bf:CtFt05} illustrates our results for the above calculations with
the CS mode frequency $\omega_0=0.1\omega_c$ and initial amplitude $\alpha_0=1.2$ at strong coupling
($\eta_0=0.5$). Surprisingly, we find that the bit-flip encoding turns out to further degrade the quantum channel.
With increasing bit redundancy in the encoding, the concurrence tends to have shorter life spans and the
teleportation fidelity drops further below the classical value, despite the non-decaying $u(t)$ at long time.
The bit-flip encoding here thus not only would not help recover the quantum channel, but would actually
further disrupt it.

In order to see the reason for the deterioration caused by bit-flip encoding, let us apply this encoding to the
cat state \eref{cat}. This yields
\begin{eqnarray}
|Q_n\rangle = \frac{1}{\sqrt{N_n}} \left( c_1|\alpha_0\rangle^{\otimes n} + c_2|-\alpha_0\rangle^{\otimes n}\right)\,,
\label{cat_n}
\end{eqnarray}
where $N_n=1 + e^{-2n|\alpha_0|^2} (c_1^*c_2+c_1c_2^*)$. Using the prescription \eref{elm_n} for the time evolution
of the density matrix $|Q_n\rangle\langle Q_n|$ and expressing the result as an operator sum
similar to \eref{k_sum}, one finds the phase-error probability after the bit-flip encoding becomes
\begin{eqnarray}
p_e^{(n)} = \frac{1}{2}\left(1-e^{-2n(|\alpha_0|^2-|\alpha_t|^2)}\right) \, .
\label{p_en}
\end{eqnarray}
It is then clear that with increasing bit redundancy $n$ in the encoding, the phase error actually increases.
In other words, the bit-flip encoding would in fact enhance effect (b) in \eref{effects} for
coherent-state qubits. Therefore, instead of reducing environment noises, the bit-flip encoding induces
additional sources for phase errors, which further corrupt the noisy channel.

\section{\label{sec:fin}Conclusions and Discussions}
In summary, we have studied the exact dynamics of optical coherent-state qubits when they are
exposed to environment noises. The environment is modeled with a collection of simple harmonic oscillators
which interact with the qubit by exchanging energies. Making use of a coherent-state path integral
formulation for this model \cite{An,Xion}, we are able to study non-perturbatively feedback effects on
the qubit dynamics due to strong environment coupling. In particular, we examine the dynamics of a noisy
quantum channel that consists of two entangled qubits which are coupled independently to their local
environments. Due to feedback from the qubit-environment interaction at strong coupling, the time evolution of the qubit
can become non-dissipative at long time. We study the concurrence and teleportation fidelity of the noisy channel and
show that, by incorporating a phase-flip error-correcting code, a robust quantum channel can be
achieved when the qubit-environment interaction is strong. As a comparison, we also consider
a bit-flip encoding scheme, which turns out to further degrade the noisy channel.

In addition to demonstrating an approach for studying the exact dynamics of coherent-state qubits subject to environment noise,
a key finding of this work is the possibility for achieving a robust quantum channel despite strong qubit-environment interactions.
This relies not only on applying appropriate error-correcting code, but also occurrence of the non-dissipative dynamics
of $u(t)$ at long time, which depends strongly on the structure of the spectral function \cite{WM12}. For instance,
for a Lorentzian spectral function, $u(t)$ would not exhibit non-dissipative dynamics in the strong coupling regime \cite{wu12}.
Therefore, in this case even when phase-flip error-correcting code is implemented for the noisy channel, its concurrence
cannot have a non-zero steady value, and its teleportation fidelity always stays below the
classical value at long time \cite{MJ}.

Although the robust quantum channel discovered in this paper has been demonstrated for a
specific initial state \eref{ctecs}, we believe that the result should be fairly general. This is because
the robust quantum channel would emerge as long as one can deal with both amplitude reduction and phase errors
due to the environment noise properly. Since the dynamics of $u(t)$ depends solely on the Hamiltonian \eref{H}, and
not on the initial state, whenever $u(t)$ attains non-zero steady value at long time, amplitude reduction would
cease to exist. For phase errors, as noted in Sec.~\ref{sec:pfc}, the phase-flip error-correcting code
can work efficiently since the error probability $p_e$ here is always less than 1/2,
irrespective of the initial state. Therefore, a robust quantum channel can survive in the long-time
limit for any initially entangled states for suitably chosen parameters (e.g.~the coupling strength). Of course,
the quality of the quantum channel would certainly depend on the specific initial state, the essence of our
conclusions should remain valid in general.

Finally, we would like to point out that it would be interesting to try to understand the deeper reason underlying the
emergence of the robust quantum channel. For instance, can we understand it in terms of the interplay
between different dynamical maps? And if so, can we extend these results to more general settings? We
hope to pursue these lines of investigation in future works.

\begin{acknowledgments}
We would like to thanks Professors Dian-Jiun Han and Pochung Chen for useful discussions. This research is
supported by the National Science Council (NSC) of Taiwan through grant no. NSC 99-2112-M-194
-009 -MY3. It is also partly supported by the Center for Theoretical Sciences, Taiwan.
\end{acknowledgments}

\appendix

\section{\label{sec:ut}The calculations for solving $u(t)$ from Eq.~\eref{ut_eom}}
In this appendix we explain briefly how to solve for $u(t)$ from \eref{ut_eom} when the spectral function
is given by \eref{Jw}. Taking the Laplace transform of \eref{ut_eom} and using the initial condition
$u(0)=1$ (we take $t_0=0$ throughout), one can obtain the Laplace transform for $u(t)$
\begin{eqnarray}
\hat{u}(z) = \frac{1}{i\omega_0+z+\hat{g}(z)} \,.
\label{uz}
\end{eqnarray}
Here $z$ is the Laplace variable and $\hat{g}$ stands for the Laplace transform
of $g(t)$ in \eref{gt}. Using \eref{Jw} in \eref{gt}, one can find easily
\begin{eqnarray}
\hat{g}(z) = -i\eta_s\tau_c \int_0^\infty dx \frac{x^s}{x-i\tau_c\,z} e^{-x} \, ,
\label{gz_1}
\end{eqnarray}
where we have denoted $\tau_c=1/\omega_c$. The $x$-integral here can then be spelled out in full using
special functions. For example, for $s=3$ one can express the $x$-integral in terms of the
exponential integral $E_1$ \cite{AB} and find
\begin{eqnarray}
\hat{g}(z) = -i\eta_s\tau_c \left[2+i\tau_c\,z-(\tau_c\,z)^2-i(\tau_c\,z)^3e^{-i\tau_c\,z}E_1(-i\tau_c\,z) \right] \, .
\label{gz}
\end{eqnarray}
Substituting \eref{gz} back into \eref{uz}, one can then write $u(t)$ as the following Bromwich integral
\begin{eqnarray}
u(t)&=&\frac{1}{2\pi i} \int_{\zeta-i\infty}^{\zeta+i\infty}\!\!dz\,\hat{u}(z)\,e^{zt}
\nonumber\\
&=&\frac{1}{2\pi i}\int_{\zeta\tau_c-i\infty}^{\zeta\tau_c+i\infty}\!\!
dz' \frac{e^{z't'}}{i(\omega_0\tau_c-2\eta_s)+(1+\eta_s)z'+i\eta_s z'^2-\eta_s z'^3 e^{-iz'}E_1(-iz')} \,
\label{Brom}
\end{eqnarray}
with $t'\equiv t/\tau_c$. Here $\zeta$ is a real number such that any pole of $\hat{u}(z)$ would lie to the left of
the contour $z=\zeta$ and we have made the change of variable $z'=z\tau_c$ going from the first to the second equation.
Since the exponential integral $E_1(z)$ has a branch cut along the negative real axis, the integral in \eref{Brom}
can thus be separated into two parts, one coming from the pole contribution and the other from the branch cut:
\begin{eqnarray}
u(t) &=& \sum \mbox{residues} \nonumber \\
&+& \frac{1}{\pi}\int_0^\infty\!\! d\omega\,
\mbox{Im}\!\left\{\frac{1}{(\omega_0\tau_c-2\eta_s)-(1+\eta_s)\omega-\eta_s\omega^2-\eta_s\omega^3 e^{-\omega}(-E_i(\omega)+i\pi)}\right\}
e^{-i\omega t'} \, .
\nonumber\\
\label{ut-supOhm}
\end{eqnarray}
Here the first term indicates a summation over all pole contributions, and the second term arises from the deformed contour
around the negative imaginary axis. In arriving at the last expression, we have used $E_1(x e^{\pm i\pi})= -E_i(x)\mp i\pi$
for $x>0$ \cite{AB}. The integral in \eref{ut-supOhm} is now a
(half-range) Fourier integral, which can be evaluated very efficiently using fast-Fourier transform techniques \cite{NR}.
We note that when $\hat{u}(z)$ has any pole, it invariably lies over the imaginary $z$-axis and gives rise to a non-decaying
term in \eref{ut-supOhm} \cite{WM12}. It is the interference between this term and the $\omega$-integral term that gives rise to
the novel time evolution of $|u(t)|$ at strong coupling (see Fig.~\ref{fig:ut}(b)).

For the sub-Ohmic case with $s=1/2$ considered in this paper, the calculation proceeds similarly to above and we find
\begin{eqnarray}
u(t) &=& \sum \mbox{residues} \nonumber \\
&+& \frac{1}{\pi}\int_0^\infty\!\! d\omega\,
\mbox{Im}\!\left\{\frac{1}{(\omega_0\tau_c-\sqrt{\pi}\eta_s)-\omega-i\pi\eta_s\sqrt{\omega}(e^{-\omega}+i\frac{2}{\sqrt{\pi}} F(\sqrt{\omega}))}\right\}
e^{-i\omega t'} \, ,
\label{ut-subOhm}
\end{eqnarray}
where $F(z)$ is Dawson's integral
\begin{eqnarray}
F(z) = e^{-z^2} \int_0^z e^{x^2} dx
\end{eqnarray}
which can again be evaluated numerically with high efficiency \cite{NR}. For Ohmic coupling, setting $s=1$ in \eref{gz_1}, one can carry
out the calculation similarly and obtain
\begin{eqnarray}
u(t) &=& \sum \mbox{residues} \nonumber \\
&+& \frac{1}{\pi}\int_0^\infty\!\! d\omega\,
\mbox{Im}\!\left\{\frac{1}{(\omega_0\tau_c-\eta_s)-\omega[1+\eta_s e^{-\omega}(-E_i(\omega)+i\pi)]}\right\}
e^{-i\omega t'} \, .
\label{ut-Ohmic}
\end{eqnarray}
The results \eref{ut-supOhm}, \eref{ut-subOhm}, and \eref{ut-Ohmic} are in complete agreement with those found in Ref.~\onlinecite{WM12}.



\begin{thebibliography}{99}
\bibitem{note}
Note that here we have used normalized coherent states, instead of those in
Refs.~\onlinecite{An} and \onlinecite{Xion}.
\bibitem{note4}
Namely, the reduction of the coherent-state amplitude from $\alpha_0$ to
$\alpha_t$. We avoid using the term ``amplitude damping" here, because for qubits
this usually refers to dissipations due to random transitions from the higher energy
basis state to the lower one \cite{NC}.
\bibitem{note2}
Since $\hat{Z}$ would in general change the norm of the state vector that it acts on
(say, the cat state \eref{cat}), we have $\hat{Z}^\dagger\hat{Z}\neq\hat{I}$.
Now that $\hat{Z}^2=\hat{I}$, it implies that $\hat{Z}^\dagger\neq\hat{Z}$.
\bibitem{note3}
In fact, for $\eta_0=0.01$ the dacay rate of $|u(t)|$ does not fully agree with that
of \eref{markov}, i.e.~$J(\omega_0)/2$. For better agreement with the Markovian result
\eref{markov}, one must consider smaller values of coupling strength, such as $\eta_0=0.001$.
Since Markovian dynamics is not our major concern here, we avoid using such small coupling strengths.
\end{thebibliography}
\end{document}